\documentclass{aastex63}

\submitjournal{ApJ}
\usepackage{threeparttablex}
\usepackage{longtable}
\usepackage{graphicx}
\usepackage{float}
\usepackage{url}
\usepackage{catchfile}
\begin{document}

\title{First Discovery of a Fast Radio Burst at 350\,MHz by the GBNCC Survey}
\correspondingauthor{E.~Parent}
\author[0000-0002-0430-6504]{E.~Parent}
\email{parente@physics.mcgill.ca}
\affiliation{Dept.~of Physics and McGill Space Institute, McGill Univ., Montr\'{e}al, QC H3A 2T8, Canada}

\author[0000-0002-3426-7606]{P.~Chawla}
\affiliation{Dept.~of Physics and McGill Space Institute, McGill Univ., Montr\'{e}al, QC H3A 2T8, Canada}

\author[0000-0001-9345-0307]{V.~M.~Kaspi}
\affiliation{Dept.~of Physics and McGill Space Institute, McGill Univ., Montr\'{e}al, QC H3A 2T8, Canada}

\author{G.~Y.~Agazie}
\affiliation{Department of Physics and Astronomy, West Virginia University, Morgantown, WV 26501}
\affiliation{Center for Gravitational Waves and Cosmology, West Virginia University, Chestnut Ridge Research Building, Morgantown, WV 26505}

\author[0000-0003-4046-884X]{H.~Blumer}
\affiliation{Department of Physics and Astronomy, West Virginia University, Morgantown, WV 26501}
\affiliation{Center for Gravitational Waves and Cosmology, West Virginia University, Chestnut Ridge Research Building, Morgantown, WV 26505}

\author[0000-0002-2185-1790]{M.~DeCesar}
\affiliation{Department of Physics, 730 High St., Lafayette College, Easton, PA 18042, USA}

\author[0000-0001-5645-5336]{W. Fiore}
\affiliation{Department of Physics and Astronomy, West Virginia University, Morgantown, WV 26501}
\affiliation{Center for Gravitational Waves and Cosmology, West Virginia University, Chestnut Ridge Research Building, Morgantown, WV 26505}

\author[0000-0001-8384-5049]{E.~Fonseca}
\affiliation{Dept.~of Physics and McGill Space Institute, McGill Univ., Montr\'{e}al, QC H3A 2T8, Canada}

\author[0000-0003-2317-1446]{J. W. T.~Hessels}
\affiliation{ASTRON, the Netherlands Institute for Radio Astronomy, Oude Hoogeveensedijk 4, 7991 PD Dwingeloo, The Netherlands}
\affiliation{Anton Pannekoek Institute for Astronomy, University of Amsterdam, Postbus 94249, 1090 GE Amsterdam, The Netherlands}

\author[0000-0001-6295-2881]{D.~L.~Kaplan}
\affiliation{Center for Gravitation, Cosmology, and Astrophysics, Department of Physics, University of Wisconsin-Milwaukee, P.O. Box 413,Milwaukee, WI 53201, USA}

\author[0000-0001-8864-7471]{V.~I.~Kondratiev}
\affiliation{ASTRON, the Netherlands Institute for Radio Astronomy, Oude Hoogeveensedijk 4, 7991 PD Dwingeloo, The Netherlands}
\affiliation{Astro Space Centre, Lebedev Physical Institute, Russian Academy of Sciences, Profsoyuznaya Str. 84/32, Moscow 117997, Russia}

\author{M.~LaRose}
\affiliation{Department of Physics and Astronomy, West Virginia University, Morgantown, WV 26501}
\affiliation{Center for Gravitational Waves and Cosmology, West Virginia University, Chestnut Ridge Research Building, Morgantown, WV 26505}

\author[0000-0002-2034-2986]{L.~Levin}
\affiliation{Jodrell Bank Centre for Astrophysics, School of Physics and Astronomy, The University of Manchester, Manchester, M13 9PL, UK}

\author[0000-0002-2972-522X]{E.~F.~Lewis}
\affiliation{Department of Physics and Astronomy, West Virginia University, Morgantown, WV 26501}
\affiliation{Center for Gravitational Waves and Cosmology, West Virginia University, Chestnut Ridge Research Building, Morgantown, WV 26505}

\author[0000-0001-5229-7430]{R.~S.~Lynch}
\affiliation{Green Bank Observatory, P.O. Box 2, Green Bank, WV 24494, USA}

\author[0000-0001-5481-7559]{A.~E.~McEwen}
\affiliation{Center for Gravitation, Cosmology, and Astrophysics, Department of Physics, University of Wisconsin-Milwaukee, P.O. Box 413,Milwaukee, WI 53201, USA}

\author[0000-0001-7697-7422]{M.~A.~McLaughlin}
\affiliation{Department of Physics and Astronomy, West Virginia University, Morgantown, WV 26501}
\affiliation{Center for Gravitational Waves and Cosmology, West Virginia University, Chestnut Ridge Research Building, Morgantown, WV 26505}

\author{M.~Mingyar}
\affiliation{Department of Physics and Astronomy, West Virginia University, Morgantown, WV 26501}
\affiliation{Center for Gravitational Waves and Cosmology, West Virginia University, Chestnut Ridge Research Building, Morgantown, WV 26505}

\author[0000-0002-4187-4981]{H.~Al Noori}
\affiliation{Department of Physics, University of California, Santa Barbara, Santa Barabara, CA 93106, USA}

\author[0000-0001-5799-9714]{S.~M.~Ransom}
\affiliation{National Radio Astronomy Observatory, 520 Edgemont Rd., Charlottesville, VA 22903, USA}

\author[0000-0002-9396-9720]{M.~S.~E.~Roberts}
\affiliation{New York University Abu Dhabi, Abu Dhabi,
UAE}
\affiliation{Eureka Scientific, Inc., 2452 Delmer St., Suite 100, Oakland, CA 94602, USA}

\author{A.~Schmiedekamp}
\affiliation{Department of Physics, The Pennsylvania State University, Ogontz Campus, Abington, Pennsylvania 19001, USA}

\author{C.~Schmiedekamp}
\affiliation{Department of Physics, The Pennsylvania State University, Ogontz Campus, Abington, Pennsylvania 19001, USA}

\author[0000-0002-7778-2990]{X.~Siemens}
\affiliation{Center for Gravitation, Cosmology, and Astrophysics, Department of Physics, University of Wisconsin-Milwaukee, P.O. Box 413,Milwaukee, WI 53201, USA}

\author[0000-0002-6730-3298]{R.~Spiewak}
\affiliation{Centre for Astrophysics and Supercomputing, Swinburne University of Technology, PO Box 218, Hawthorn, VIC 3122, Australia}

\author[0000-0001-9784-8670]{I.~H.~Stairs}
\affiliation{Dept. of Physics and Astronomy, University of British Columbia, 6224 Agricultural Road, Vancouver, BC V6T 1Z1 Canada}

\author[0000-0002-9507-6985]{M. Surnis}
\affiliation{Department of Physics and Astronomy, West Virginia University, Morgantown, WV 26501}
\affiliation{Center for Gravitational Waves and Cosmology, West Virginia University, Chestnut Ridge Research Building, Morgantown, WV 26505}

\author[0000-0002-1075-3837]{J.~Swiggum}
\affiliation{Department of Physics, Lafayette College, Easton, PA 18042, USA}

\author[0000-0001-8503-6958]{J.~van Leeuwen}
\affiliation{ASTRON, the Netherlands Institute for Radio Astronomy, Oude Hoogeveesedijk 4,7991 PD Dwingeloo, The Netherlands}
\affiliation{Anton Pannekoek Institute, University of Amsterdam, Postbus 94249, 1090 GE Amsterdam, The Netherlands}

\begin{abstract}
We report the first discovery of a fast radio burst (FRB), FRB 20200125A, by the Green Bank Northern Celestial Cap (GBNCC) Pulsar Survey conducted with the Green Bank Telescope at 350\,MHz. 
FRB 20200125A was detected at a Galactic latitude of 58.43$^\circ$ with a dispersion measure of 179\,pc\,cm$^{−3}$, while electron density
models predict a maximum Galactic contribution of 25\,pc\,cm$^{−3}$ along this line of sight.
Moreover, no apparent Galactic foreground sources of ionized gas that could account for the excess DM are visible in multi-wavelength surveys of this region. This argues that the source is  extragalactic. The maximum redshift for the host galaxy is $z_{max}=0.17$, corresponding to a maximum comoving distance of approximately 750\,Mpc. The measured peak flux density for FRB 20200125A is 0.37\,Jy, and we measure a pulse width of 3.7\,ms, consistent with the distribution of FRB widths observed at higher frequencies.  Based on this detection and assuming an Euclidean flux density distribution of FRBs, we calculate an all-sky rate at 350\,MHz of $3.4^{+15.4}_{-3.3} \times 10^3$ FRBs sky$^{-1}$ day$^{-1}$ above a peak flux density of 0.42\,Jy for an unscattered pulse having an intrinsic width of 5\,ms, consistent with rates reported at higher frequencies. 
Given the recent improvements in our single-pulse search pipeline, we also revisit the GBNCC survey sensitivity to various burst properties. Finally, we find no evidence of interstellar scattering in FRB 20200125A, adding to the growing evidence that some FRBs have circumburst environments where free-free absorption and scattering are not significant.
\end{abstract}

\section{Introduction} \label{sec:intro}
Fast radio bursts (FRBs) are energetic, millisecond-duration radio transients of extragalactic origin (see, e.g., \citealt{phl+19} for a review). More than one hundred FRB sources have been published to date, including 20 repeating sources \citep{sshb+16,chime_rn1,kso+19,chime_rn2}, yet their origin remains an open question. Numerous models involving a diversity of progenitors and emission processes have been postulated to explain the FRB phenomenology (for a summary of proposed theories\footnote{\url{https://frbtheorycat.org}}, see \citealt{pww+19}). Some of the leading models invoke magnetars as the source of FRBs. Recently, two radio bursts having luminosities comparable to FRBs were detected at a dispersion measure (DM) and position consistent with the Galactic magnetar SGR 1935+2154 \citep{s+chime20,bkr+20,zjm+20,ksj+20} during a known active phase of X-ray emission \citep{bbd+20,p20,f+fermi20,yge+20}. These findings support the notion that extragalactic magnetars are a source of at least some FRBs.

Observations of FRBs at multiple radio frequencies can help constrain proposed emission mechanisms as well as their local environments based on measurements of properties arising from propagation effects. Initially, most FRBs were detected at a frequency of $\sim$ 1.4 GHz. Follow-up observations of the original repeating FRB 121102 \citep{sshb+16} in the 4$-$8 GHz range by \cite{msh+18} and \cite{gsp+18} showed that FRBs can emit at higher frequencies. Over the past two years, many FRBs have been detected by the Canadian Hydrogen Intensity Mapping Experiment Fast Radio Burst Project (CHIME/FRB; \citealt{chime18}) down to frequencies of 400\,MHz \citep{chime19,jcf+19}. Thus, one must conclude that the environment of some FRBs is optically thin to these frequencies, and any frequency cutoff or turnover invoked in proposed mechanisms must occur below these frequencies (e.g., \citealt{rl19}). 

A particularly interesting discovery reported by the CHIME/FRB Collaboration is the repeating source FRB 180916.J0158+65 which shows a 16.35-day modulation in its burst activity \citep{chime_periodic_frb}. In Spring 2020, follow-up observations of FRB 180916.J0158+65 in the 300$-$400\,MHz frequency range \citep{cab+20,pbp+20} provided the lowest-frequency detections of FRBs to date. 
These targeted observations provide the only detections of FRBs below 400\,MHz in spite of previous efforts to survey the radio sky for FRBs in the low-frequency regime (e.g., \citealt{cvh+14,kca+15,ttw+15,rbm+16,dsm+16,ckj+17,tef+19,rms+20}). 
These new results, in addition to the numerous unscattered events reported by the CHIME/FRB Collaboration where the emission is detectable throughout their observing band, suggest that new FRBs will be detected by such large-scale programs.

\cite{ckj+17} reported on a search for FRBs with the ongoing Green Bank Northern Celestial Cap (GBNCC) Pulsar Survey \citep{slr+14}, conducted with the 100-m Robert C. Byrd Telescope at the Green Bank Observatory (GBT) at 350\,MHz. They searched 65,000 GBNCC sky pointings collected between 2009 and 2016 for FRBs, totalling 84 days of on-sky time, and found no FRB signals with S/N values greater than 10. From their non-detection, \cite{ckj+17} calculated a 95\% confidence upper limit on the FRB rate\footnote{For an unscattered pulse width of 5 ms} of 3.6$\times 10^3$ FRBs sky$^{-1}$ day $^{-1}$ above a peak flux density of 0.63\,Jy at 350\,MHz. Since the time of this study, $\sim$33,000 additional GBNCC pointings have been searched for fast transients. 

Here, we report the first FRB discovered by GBNCC, FRB 20200125A, which is also the first FRB discovery to emerge from a large-scale, non-targeted survey below 400\,MHz. 
In Section \ref{sec:obs}, we describe the survey and data processing, and discuss the discovery of FRB 20200125A and compare the significance of the signal to other GBNCC confirmed discoveries of Rotating Radio Transients (RRATs). Burst properties are provided in Section \ref{sec:burst}. In Section \ref{sec:extragal}, we discuss the evidence for the source being extragalactic. An all-sky FRB rate calculation is presented in Section \ref{sec:rate}. In the latter, we additionally revisit the sensitivity of the survey to single-pulse events. 
Our results are summarized in Section \ref{sec:conclu}. 
\vspace{0.9cm}

\section{Observations \& Discovery} \label{sec:obs}
\subsection{Survey Description and Data Processing} \label{sec:survey}
An all-sky GBNCC survey for pulsars and dispersed radio pulses began in 2009 with the GBT at a central observing frequency of 350\,MHz. Data are sampled every 81.92\,$\mu$s and recorded in 4096 frequency channels spanning 100\,MHz with the Green Bank Ultimate Pulsar Processing Instrument\footnote{\url{https://safe.nrao.edu/wiki/bin/view/CICADA/GUPPiUsersGuide}} (GUPPI). All survey pointings consist of 120-s integrations on a particular sky location with a full width at half-maximum (FWHM) beam size of 36$'$. A total of 125,000 sky pointings are necessary to cover the entire sky visible by GBT ($\delta > -40^{\circ}$), 90\% of which have been observed as of 2020 July.

Data are processed with a \texttt{PRESTO}-based\footnote{\url{https://www.cv.nrao.edu/~sransom/presto/}} \citep{r01} pipeline that is run on the B\'eluga cluster\footnote{\url{https://docs.computecanada.ca/wiki/Beluga}} operated by Calcul Qu\'ebec and Compute Canada in Montr\'eal, Canada. Following of the excision of radio frequency interference (RFI), pointings are dedispersed at various DM trials up to 3000\,pc\,cm$^{-3}$. We note however that prior to 2013, data were only searched up to a maximum DM of 500\,pc\,cm$^{-3}$. In total, about 21\% of the $\sim$112,000 pointings collected and processed as of 2020 May have only been searched up to 500\,pc\,cm$^{-3}$, and will be reprocessed in the future. 

Time series are searched for periodic signals with both an FFT-based search and a Fast-Folding Algorithm \citep{pkr+18}, and single pulses are searched with \texttt{PRESTO}'s \texttt{single\_pulse\_search.py}. The latter is based on a matched-filtering algorithm that convolves the time series with boxcars having different widths. Single-pulse events having widths $<$ 100\,ms detected above a signal-to-noise ratio (S/N) of 5 are saved for further analysis. Once all time series have been searched, events are sifted, grouped and ranked with the \texttt{RRATtrap} grouping algorithm \citep{kkl+15}. \cite{ckj+17} describe in detail the grouping component of the GBNCC pipeline, and a more general description of the survey and data processing is provided by \cite{slr+14}.

\begin{figure*}[ht]
  \centering
    \includegraphics[width=0.90\textwidth]{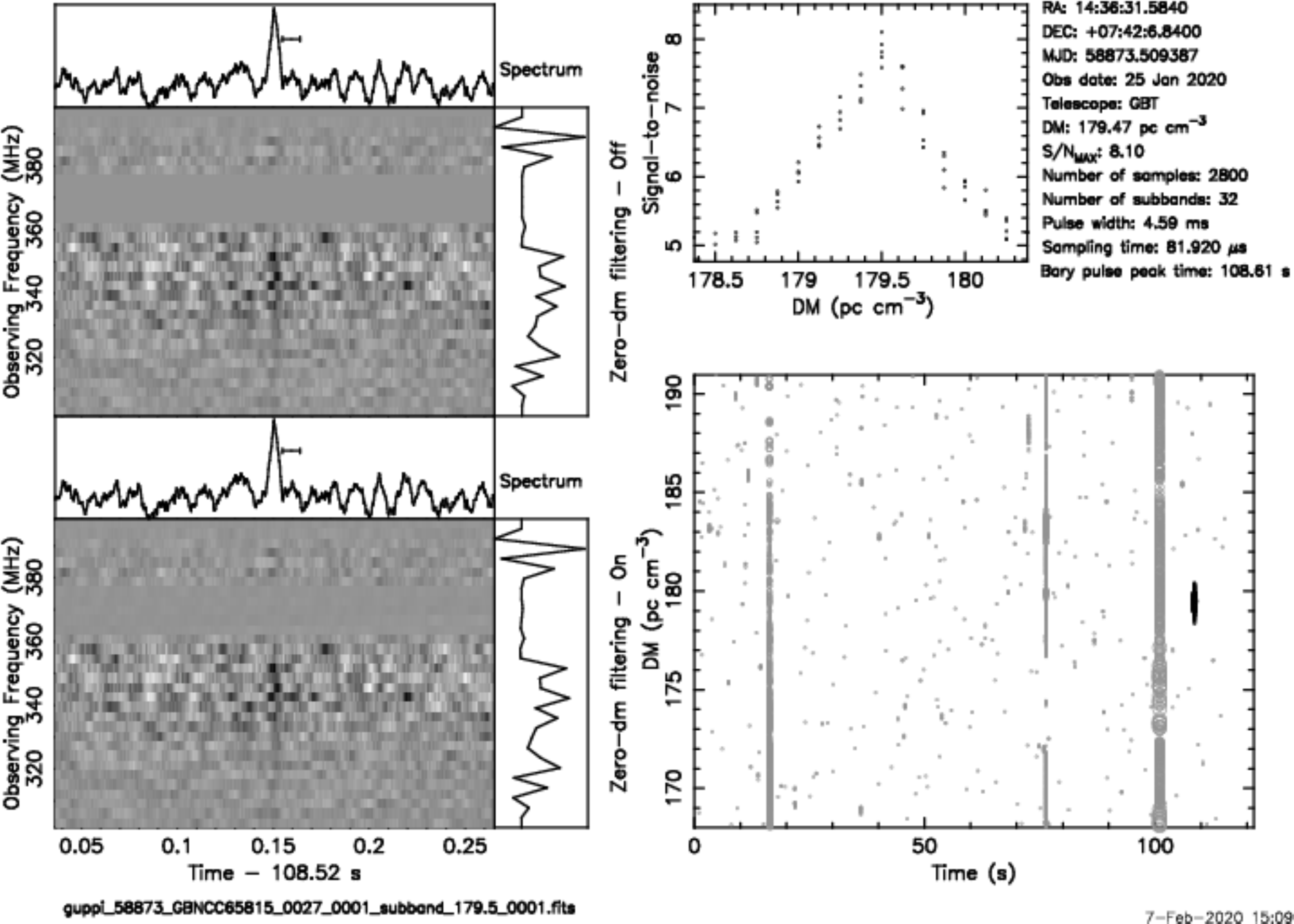}
    \caption{Discovery SPD plot for FRB 20200125A. \textit{Left panels:} Dedispersed dynamic spectra without (\textit{top}) and with (\textit{bottom}) zero-DM filtering \citep{ekl+09}. Data in the 360$-$380\,MHz frequency range are masked as a result of persistent RFI at the telescope site. Dedispersed time series, shown above their respective dedispersed dynamic spectra, are produced by summing the frequency channels below. The boxcar width of the highest-S/N event in the group is shown as a horizontal bar next to the pulse. Spectra are shown on the right of each dynamic spectrum. \textit{Top right panel:} S/N vs DM of grouped events (black dots in the bottom right panel). The peak of the curve corresponds to the optimal DM for the single-pulse candidate. Header and candidate information are displayed on the right of the S/N vs DM plot. \textit{Bottom right panel:} DM of single pulse detections as a function of time, where the size of the dots scales with S/N. Single pulse events selected by the grouping algorithm that are associated to FRB 20200125A are shown in black, while all other events for that DM range are in grey. Vertical bars spanning a large DM range (e.g. at Time\,=\,100\,s) are sporadic RFI events.   
    }
    \label{fig:discovery}
\end{figure*}

\begin{figure}[ht]
  \centering
    \includegraphics[width=0.55\textwidth]{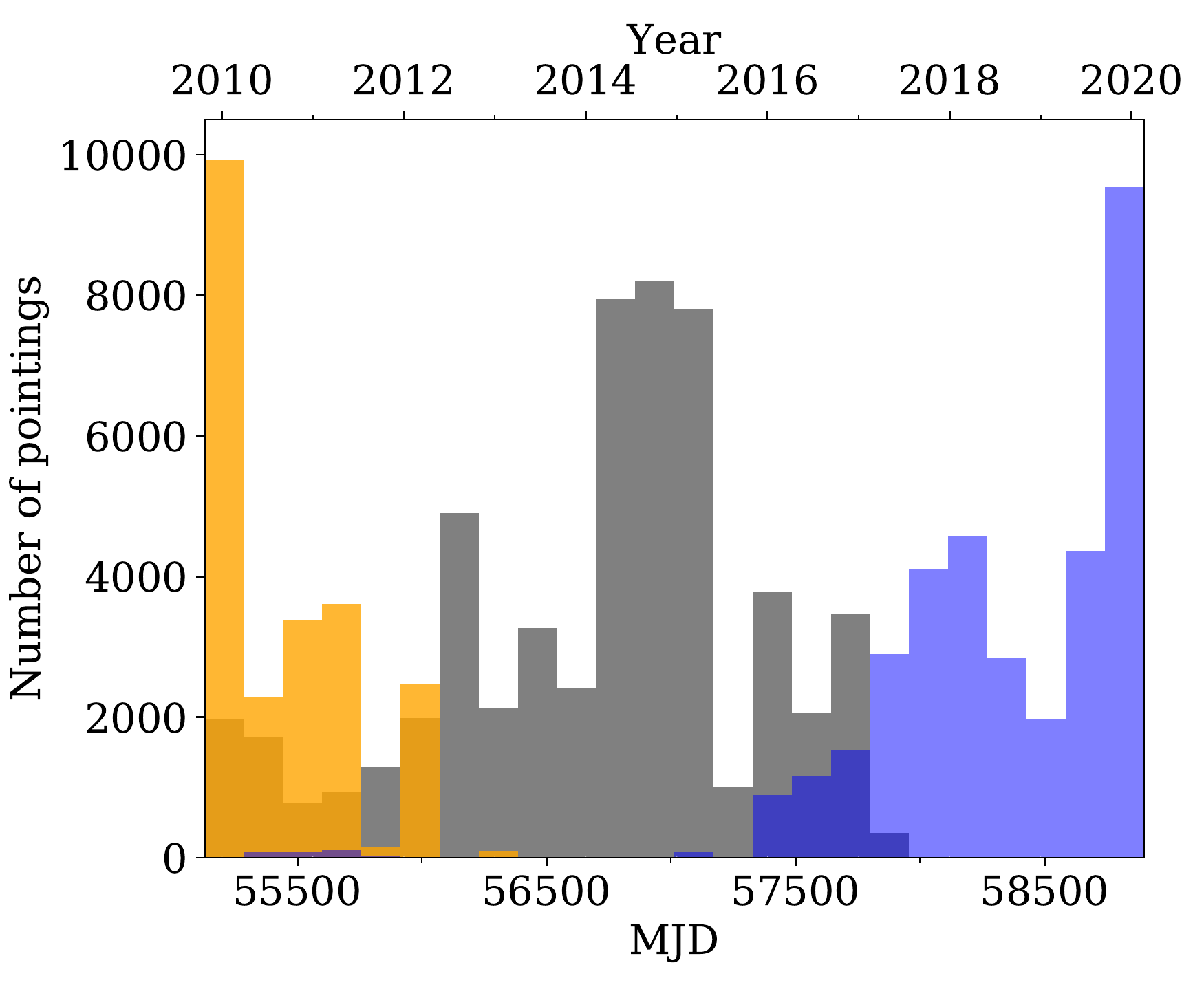}
    \caption{Distribution of GBNCC pointings searched for single pulses based on the pipeline version used for the search as a function of their observation date. Pointings shown in orange were searched only up to a DM of 500\,pc\,cm$^{-3}$ while those in grey were searched up to a DM of 3000\,pc \,cm$^{-3}$. Pointings in blue were also searched up to a DM of 3000\,pc\,cm$^{-3}$, but were processed through our upgraded single-pulse pipeline (as described in Section \ref{sec:survey}) and had their processing output visually inspected via the CyberSKA platform. 
    }
    \label{fig:processing}
\end{figure}

In spring of 2017, upgrades to the single-pulse searching component of the pipeline were implemented. Single-pulse diagnostic plots (``SPD" plot, see e.g. Figure \ref{fig:discovery}) are now automatically generated for all single-pulse candidates, which are groups that are ranked as being potentially astrophysical by \texttt{RRATtrap} (i.e., not ranked as noise and/or RFI). SPD plots are then uploaded to the CyberSKA online portal \citep{kab+11} for visual inspection and classification. On average, between 25 and 30 candidates per pointing are produced. Approximately 33,000 pointings have been processed with this upgraded pipeline, representing $\sim$29\% of all observed pointings. In Figure \ref{fig:processing} we show the temporal distribution of GBNCC pointings processed through the different versions of the search pipeline.

Nearly one million single-pulse candidates have been generated by this new pipeline, 6\% of which have been visually inspected on CyberSKA. The remaining 94\% either have S/N$\,<\,$8 and/or pulse widths $>\,$50\,ms, and are therefore likely to be RFI or noise. 
We have developed various diagnostic ratings that are computed based on the candidates attributes, and those are used by CyberSKA users to filter out obvious false positives caused by RFI, which makes up the large majority of candidates. 
Among classified candidates are eight new RRATs\footnote{Basic properties can be found on GBNCC's discovery page, \url{http://astro.phys.wvu.edu/GBNCC/}. Timing properties will be reported in a future publication.} as well as the event we present in this work, FRB 20200125A.  

\subsection{Discovery of FRB 20200125A} \label{sec:discovery}
FRB 20200125A was identified while inspecting candidates produced by the processing pipeline on the CyberSKA portal. The burst, shown in Figure \ref{fig:discovery}, was detected on 2020 January 25 
12:15:19.61 UTC (at 350\,MHz) in a GBNCC beam pointed at the sky position $l\,=\,359.83^\circ$ and $b\,=\,58.43^\circ$. Our pipeline identified the burst at a DM of 179\,$\pm$\,2\,pc\,cm$^{-3}$ with a S/N\,=\,8.1 and best-matched boxcar width of 4.59\,ms. Optimized parameters are presented in Section \ref{sec:burst}. 

Along this line of sight, the maximum Galactic DM predicted by the NE2001 \citep{ne2001} and YMW16 \citep{ymw16} electron density models are 25\,pc\,cm$^{-3}$ and 24\,pc\,cm$^{-3}$, respectively, far less than the DM of this candidate FRB even when considering a reasonable model uncertainty (typically $\sim 25\%$). We note, however, that these models only account for the interstellar medium and do not include the Milky Way halo, which could contribute another 50--80,pc\,cm$^{-3}$ to the FRB DM \cite{pz19}. Even in the most conservative scenario, the combined contribution of the Galactic halo and interstellar medium do not explain the anomalous amount of DM observed in FRB 20200125A.  
We therefore classify this event as extragalactic, and further discuss arguments disfavoring a Galactic origin in Section \ref{sec:extragal}. Considering the marginal S/N detection of the candidate, we first evaluate its significance and likelihood of being astrophysical.     

\subsection{Candidate significance} \label{sec:cand}
If the noise were normally distributed, it would be highly improbable for a false-positive detection to occur with S/N\,=\,8.1. The RFI that is present, however, exhibits a variety of temporal and spectral features causing the reliability of statistical probabilities determined assuming Gaussian statistics to be compromised. Assessing the significance of an event based on its S/N is therefore not trivial. 

As shown the bottom right panel in Figure \ref{fig:discovery}, significant RFI remained present in the beam despite having applied standard RFI-mitigation routines. Such persistent interference increases the time series root mean square (rms) and introduces variation in the baseline level, which reduces the S/N calculated by \texttt{single\_pulse\_search.py}. 

Using the same S/N metric, we calculate a corrected S/N that is more representative of the statistical noise around the time of the burst (i.e., excluding samples corrupted by the sporadic RFI visible in Figure \ref{fig:discovery}). We produced a shortened time series at the best-DM of the burst but only processed data within 3\,s of the time of the burst, where we expect a negligible amount of persistent RFI. Running the \texttt{single\_pulse\_search.py} code on this time series increased the S/N of the burst to 8.4 for the same boxcar width.     

To assess the probability that our candidate FRB 20200125A is real, we examine the S/N distribution of all candidates uploaded to the CyberSKA platform that were manually classified upon inspection as being either 1) potentially astrophysical, 2) detection of a known source, or 3) RFI/noise. This approach was used by \cite{pab+18} for assessing the significance of their FRB 141113 candidate, discovered by the PALFA survey \citep{cfl+06} at a similarly marginal S/N value. For completeness, we visually inspected and classified all GBNCC candidates having S/Ns\,$\geq$\,8 and pulse widths $\leq\,50\,$ms.

\begin{figure*}[ht]
  \centering
    \includegraphics[width=0.90\textwidth]{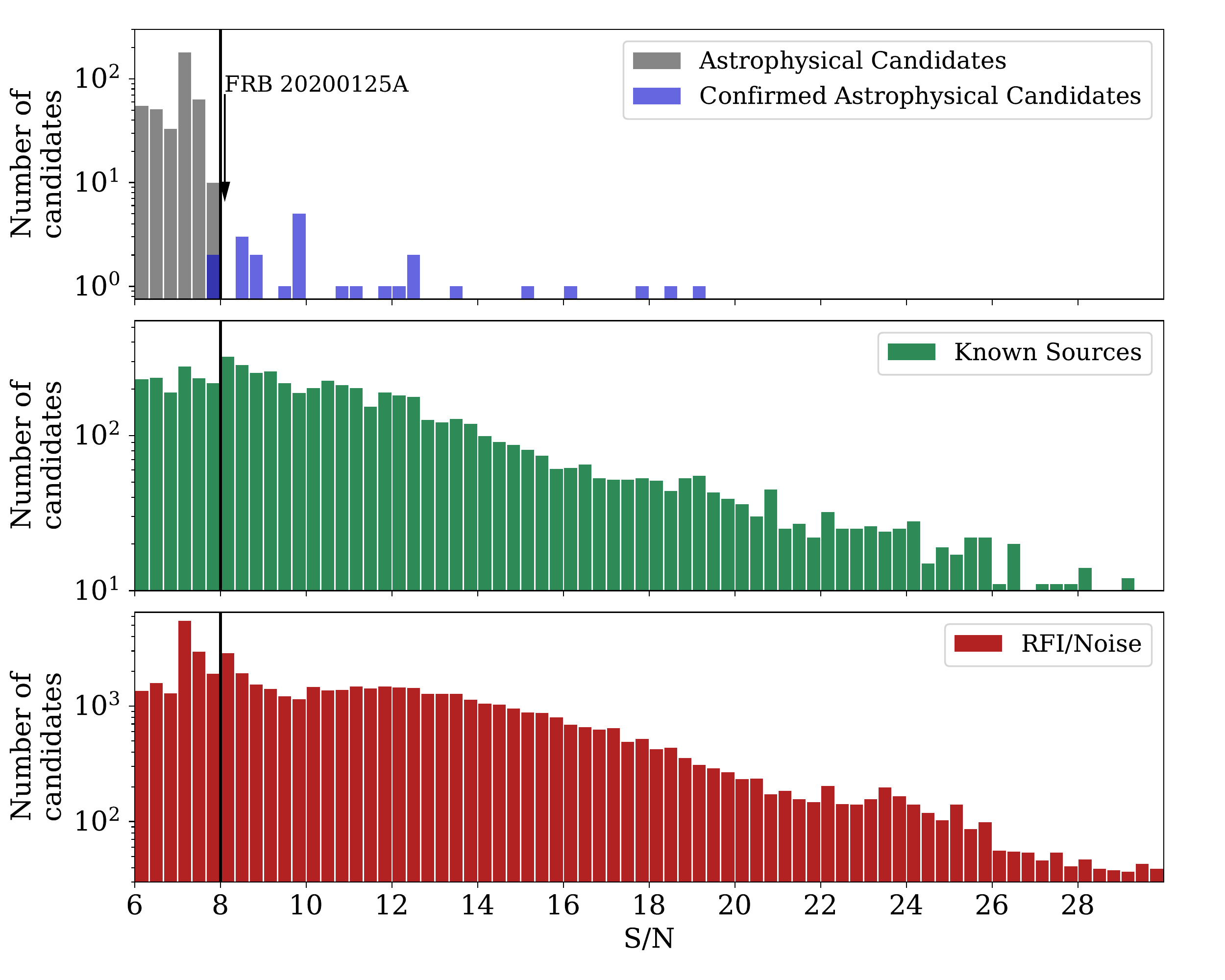}
    \caption{S/N distribution of all single-pulse candidates visually classified by members of the GBNCC collaboration on the CyberSKA platform, with the exception of some confirmed discoveries made by the survey prior to our pipeline upgrades. \textit{Top panel:} Candidates that were classified as being potentially astrophysical (grey) and confirmed discoveries (blue), which represents $<1\%$ of all manually classified candidates. The discovery-S/N of FRB 20200125A is shown by the black arrow. \textit{Middle panel:} Detections of known sources, representing $\sim$11\% of all classified candidates. \textit{Bottom panel:} Candidates classified as RFI and/or noise, which account for the remaining $\sim$88\% of all classifications. We note that not all candidates on CyberSKA have been classified: the distributions to the left of the black vertical lines (i.e., having S/N$\,<\,$8) are incomplete and therefore do not represent the true distribution of all candidates generated by the pipeline. 
    }
    \label{fig:snr_histo}
\end{figure*}

In Figure \ref{fig:snr_histo}, we show the S/N distribution of the 60,000 classified candidates based on their qualitative attributes. We see that candidates classified as re-detections of known sources (middle panel) and RFI (bottom panel) have relatively flat S/N distributions and their S/Ns extend to large values, while there is a clear shortage of candidates classified as likely astrophysical sources (top panel) for S/Ns $\gtrsim$ 8. Promising astrophysical candidates with S/N values as low as 7.7 have been confirmed as astrophysical through follow-up observations. Most importantly, \textit{all} astrophysical candidates with S/N $>$ 8 have been confirmed to be real pulsars or RRATs. Consequently, we henceforth argue that FRB 20200125A is a genuine cosmic event. 

\section{Burst properties} \label{sec:burst}
In order to characterize the burst and confidently distinguish real emission from noise, we downsampled the data to a resolution of 655.36\,$\mu$s. Reducing the temporal resolution is further justified by the minimum broadening resulting from intra-channel dispersive smearing at the burst's DM, which is estimated as 841\,$\mu$s at 350\,MHz. 

A variety of techniques have been used to perform DM-optimizations in recent FRB studies. For high-S/N, broadband-emitting bursts that do not exhibit a complex pulse morphology, times of arrival (TOAs) can be extracted from multiple frequency sub-bands and a DM-fit of the TOAs can be done following standard pulsar timing techniques \citep[e.g.,][]{sch+14,sshc+16}. Metrics that optimize the pulse structure have been developed for determining the DM of bursts having complex time-frequency structures, such as the sub-burst downward frequency drifts observed in repeating FRBs \citep[e.g.,][]{hss+19,jcf+19,chime19}.
Given the faintness of FRB 20200125A, we determine a best-fit DM using only a standard S/N-optimization approach. We obtain a DM value of 179.47\,$\pm$\,0.05\,pc\,cm$^{-3}$, in agreement with the initial pipeline detection.
We then corrected for the receiver bandpass by calculating the system time-averaged rms noise in each frequency channel, which were converted into flux density units using the radiometer equation:  
\begin{equation}
    \Delta S_{i} = \frac{\beta\left(T_{rcv} + T_{sky}\right)}{G \sigma_i \sqrt{n_p t_s \Delta\nu_i}},
\end{equation}
where $\Delta S_{i}$ is a scaling factor that corrects for the system equivalent flux density of channel $i$ in Jy, $\beta$ is the digitization factor, $T_{rcv}$ is the receiver temperature, $T_{sky}$ is the sky temperature, $G$ is the telescope gain, $\sigma_i$ is the rms fluctuation in channel $i$, $n_p$ is the number of polarizations summed, $t_s$ is the temporal resolution and $\Delta\nu_i$ is the width of channel $i$. Our data 8-bit data have a digitization factor $\beta\,=\,1.1$, and for the GBT 350-MHz receiver, $T_{rcv}\,=\,23\,$K, $G\,=\,2\,$K\,Jy$^{-1}$ and $n_p\,=\,2$.
Assuming a spectral index\footnote{The spectral index $\alpha$ is defined as $S_{\nu}\propto\nu^{\alpha}$, where $S_{\nu}$ is the flux density and $\nu$ is the observing frequency.} $\alpha$ of $-$2.55 \citep{hss+82}, we calculated the sky temperature at the frequency of each channel ($T_{sky}$ = 40.7\,K at the central frequency 350\,MHz) using the 408-MHz all-sky map from \cite{rdb+15}.

For each frequency channel, we subtracted the off-pulse mean and normalized the data with the off-pulse $\Delta S_{i}$ values. Resulting receiver gain fluctuations, notably in response to strong RFI, are not taken into account in the above estimation. Therefore, we performed the calibration considering only a 100-ms window around the burst time, and produced a band-averaged time series by summing all calibrated frequency channels. 

To ensure that the peak flux density and fluence are measured as robustly as possible, we randomly selected five 100-ms segments from the same pointing that do not contain the FRB signal. Realizations of calibrated data were then generated from these segments following the procedure described above.
We then inserted the on-pulse data into the calibrated segments to produce different noise realizations and better evaluate the uncertainties on the burst peak flux density and fluence. The peak flux density of the FRB was estimated as the highest value in the band-averaged time series. Fluence estimates were obtained by integrating the time series segments. We adopt the mean and standard deviation of these realizations as our final measurements. The measured peak flux density and fluence are 0.368$\,\pm\,$0.012\,Jy and 1.26$\,\pm\,$0.08\,Jy\,ms, respectively. We fit a Gaussian model to the burst to determine the pulse FWHM and used the best-fit mean as the burst peak time (see Table \ref{tab:burst}). 

Multipath scattering of signals propagating in an inhomogeneous ionized medium leads to frequency-dependent time delays. Such scattering-induced broadening is observable as a one-sided exponential tail in a pulse profile characterized by a scattering timescale $\tau_s \propto \nu^{-4}$. Using the convolution of a Gaussian profile and a one-sided exponential decay to describe the pulse broadening \citep{m14}, we attempt to measure a scattering timescale for FRB 20200125A and fit the above model to the band-averaged time series. Our best-fit model is consistent with a negligible exponential tail having a characteristic $\tau_s$ of $50\,\mu$s, shorter than the data resolution. Additionally, we measure the width of pulse profiles produced by separating the total bandwidth into three 33-MHz subbands, each dedispersed at the best-fit DM we obtained for the entire band, and summing them individually into band-averaged time series. We fit both the above pulse-broadening function and a single Gaussian to the pulse profiles.  Profiles at different frequencies have best-fit pulse widths consistent with the measurement reported in Table \ref{tab:burst}. Results from both approaches suggest that the presence of scattering, if any, is temporally unresolved. We also note that NE2001 predicts a scatter-broadening smaller than $9\,\mu$s along the line of sight to FRB 20200125A, while YMW16 predicts an even smaller timescale $<\,1\,\mu$s. We therefore adopt the temporal resolution of the calibrated data as an upper limit on the pulse broadening caused by scattering, i.e. $\tau_s < 655.36\,\mu$s. We recognize however that the low S/N of FRB 20200125A limits our ability to accurately measure a scattering timescale. 

Because GBNCC survey pointings are recorded in incoherent filterbank mode, only total intensity data are available. Polarization properties cannot be determined from this  detection.
\setlength{\tabcolsep}{0.5mm}
\LTcapwidth=1.\linewidth
\newcommand{\longtableone}{\hline
\multicolumn{1}{l}{Parameter} & \multicolumn{1}{c}{Measured value}  \\ \hline 
R.A.$^a$ (J2000)                    & 14:36:31.58 \\ 
Dec.$^a$ (J2000)                    & +07:42:06.84 \\
$l^a$ (deg)                         & 359.83    \\
$b^a$ (deg)                         & 58.43     \\
Event time$^b$ (MJD)                & 58873.510643634 \\ 
DM (pc\,cm$^{-3}$)                  & 179.47(5)  \\
Width$^c$ (ms)                      &  3.7(4) \\
Peak Flux density  (Jy)             & 0.368(12) \\
Fluence (Jy ms)                     & 1.26(8) \\
Scattering timescale ($\mu$s)       & $<$ 655.36 \\
\hline
\multicolumn{1}{l}{Parameter} & \multicolumn{1}{c}{Derived value}  \\ 
\hline 
DM$_{NE2001}^d$ (pc\,cm$^{-3}$)           & 25  \\
DM$_{YMW16}^d$ (pc\,cm$^{-3}$)            & 24  \\
Maximum redshift, $z_{max}$               & 0.17        \\
Maximum comoving distance, $D_{max}$ (Mpc) & 750       \\
\hline 
\insertTableNotes }%
\begin{ThreePartTable}
\begin{TableNotes}[flushleft]
\footnotesize
\item [$a$] Position of the beam center, which has a FWHM of 36$'$. 
\item [$b$] Barycentric time measured at 350\,MHz.
\item [$c$] Full width at half maximum of a single-Gaussian fit to the profile.
\item [$d$] Model-dependent maximum Galactic DM contribution predicted along the line of sight.
\end{TableNotes}
\begin{longtable}{p{6.25cm} c }
    \caption{\normalsize Measured and inferred properties of FRB 20200125A.}
    \label{tab:burst} \\ 
    \hline
\multicolumn{1}{l}{Parameter} & \multicolumn{1}{c}{Measured value}  \\ \hline 
R.A.$^a$ (J2000)                    & 14:36:31.58 \\ 
Dec.$^a$ (J2000)                    & +07:42:06.84 \\
$l^a$ (deg)                         & 359.83    \\
$b^a$ (deg)                         & 58.43     \\
Event time$^b$ (MJD)                & 58873.510643634 \\ 
DM (pc\,cm$^{-3}$)                  & 179.47(5)  \\
Width$^c$ (ms)                      &  3.7(4) \\
Peak Flux density  (Jy)             & 0.368(12) \\
Fluence (Jy ms)                     & 1.26(8) \\
Scattering timescale ($\mu$s)       & $<$ 655.36 \\
\hline
\multicolumn{1}{l}{Parameter} & \multicolumn{1}{c}{Derived value}  \\ 
\hline 
DM$_{NE2001}^d$ (pc\,cm$^{-3}$)           & 25  \\
DM$_{YMW16}^d$ (pc\,cm$^{-3}$)            & 24  \\
Maximum redshift, $z_{max}$               & 0.17        \\
Maximum comoving distance, $D_{max}$ (Mpc) & 750       \\
\hline 
\insertTableNotes 
\end{longtable} 
\end{ThreePartTable}

\section{Galactic or Extragalactic?} \label{sec:extragal}

We now consider whether FRB 20200125A could be Galactic. The NE2001 and YMW16 models predict a maximum Galactic DM along the line of sight of 25 and 24\,pc\,cm$^{-3}$, respectively. Considering the $\sim 25\%$ uncertainty generally assumed for these predictions, these values are in agreement with each other. However, DM-predicted distances can be incorrect by factors of a few (e.g., \citealt{dgb+19}). This is particularly true for high-latitude pulsars where distance predictions are sensitive to small DM variations. 

We can still evaluate how likely it is that both models are underpredicting the maximum Galactic DM in the direction of FRB 20200125A by examining their performance for pulsars in that region of the sky.

We selected the 44 pulsars from the ATNF catalog\footnote{ATNF Pulsar Catalogue version 1.63, \url{https://www.atnf.csiro.au/research/pulsar/psrcat/}} \citep{mht+05} located within 25$^\circ$ of the beam center position, and looked at their distances and DMs distribution. These sources have Galactic latitude, $b$, in the range $34^\circ < b < 80^\circ$, while their DMs vary from 3.3\,pc\,cm$^{-3}$ to a maximum of 37.4\,pc\,cm$^{-3}$, $\sim$5 times smaller than that of FRB 20200125A. Ten of these 44 pulsars are in globular clusters having known distances (five in M5, four in M3 and one in M53) and seven have parallax distance measurements. 

\begin{figure}[t]
  \centering
    \includegraphics[width=0.45\textwidth]{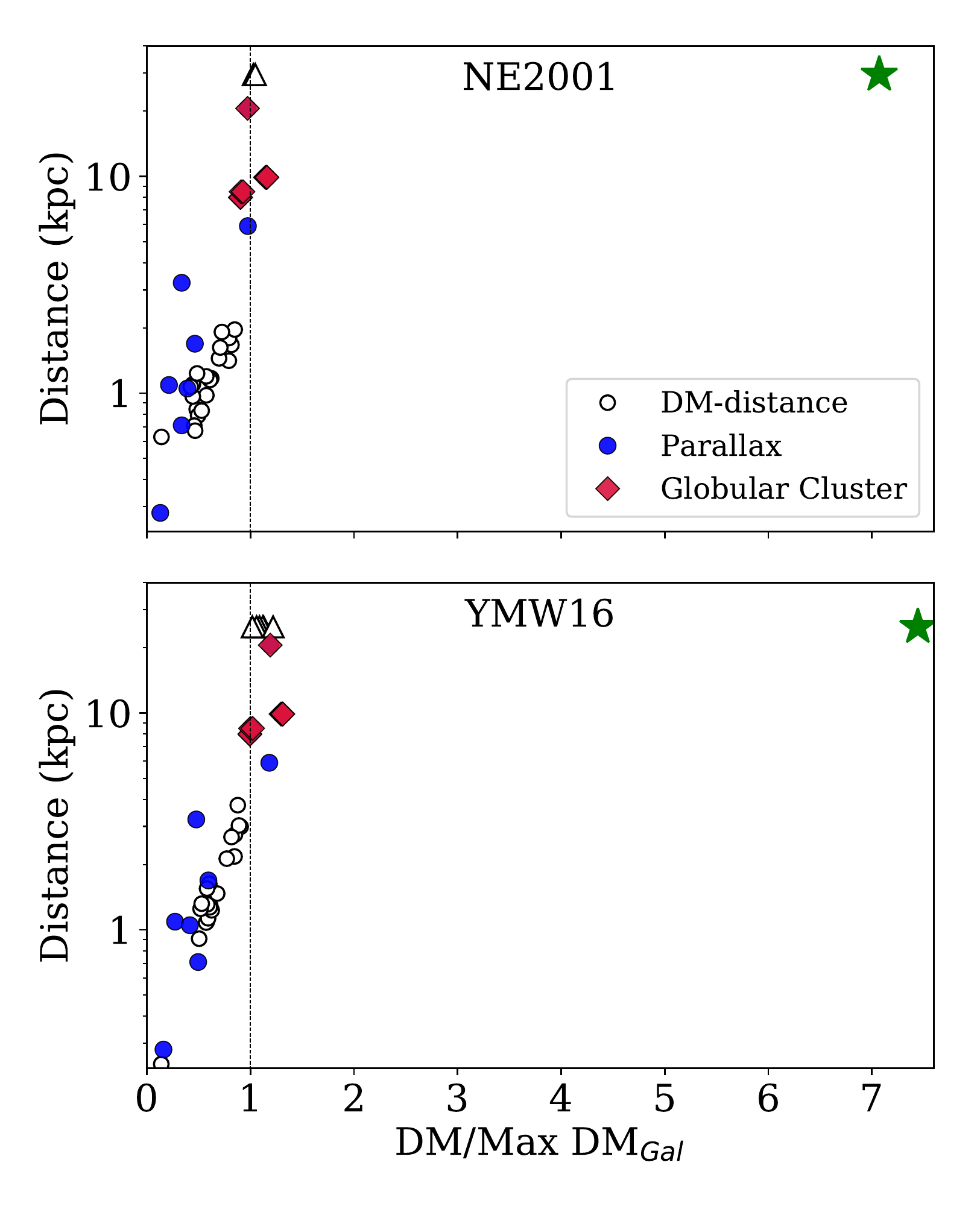}
    \caption{ Distances of pulsars within 25$^\circ$ of the beam center position in which FRB 20200125A was detected, versus the ratio of the pulsar DMs to the maximum Galactic DM along their line of sight predicted by the NE2001 electron density model (\textit{top panel}) and YMW16 (\textit{bottom panel}). The vertical dashed lines represents a DM to maximum Galactic DM ratio of unity, i.e., the predicted edge of the Milky Way. Blue-filled circles are pulsars with measured parallax distances, while red diamonds are those in globular clusters  (i.e. have known distances). Pulsars for which only DM-predicted distances are available are shown with open circles, and those having DM greater than the maximum DM predicted by electron density models are identified with upper triangle. FRB 20200125A is shown with a green star. }
    \label{fig:dm_dist}
\end{figure}

In Figure \ref{fig:dm_dist}, we show the pulsar distances against the ratio of their measured DM values to the maximum Galactic DM ($\rm{DM}/\rm{Max\,DM}_{\rm{Gal}}$) predicted by NE2001 (top panel) and YMW16 (bottom panel) along their respective lines of sight. Ratios exceeding unity imply that the electron density model places these sources beyond the Galaxy. 

Only 6/44 pulsars have DMs exceeding the maximum Galactic DM predicted by NE2001, four of which are located within the globular cluster M3 \citep{hrs+07}. On the other hand, 13/44 pulsars have DMs exceeding the maximum Galactic DM predicted by YMW16, seven of which have independent distance measurements (all but one being in globular clusters). Taking into account model uncertainties, these predictions are still consistent with a Galactic origin.
Pulsars located in M3 yield the largest ratios (1.15 and 1.31 for NE2001 and YMW16, respectively), and their corresponding DM excesses are only $\sim$ 5\,pc\,cm$^{-3}$). We therefore conclude that there is no evidence that the models severely underpredict the maximum Galactic DM in that region of the sky. 

With a DM of 179.47\,pc\,cm$^{-3}$ ($\rm{DM}/\rm{Max\,DM}_{\rm{Gal}}$ of 7.1 and 7.4 for NE2001 and YMW16, respectively), it is clear that FRB 20200125A is an outlier, regardless of the accuracy of the models. Including a contribution from the Galactic halo, which could be as high as $\sim$\,80\,pc\,cm$^{-3}$ \citep{pz19}, would still leave an  anomalous DM $>$\,75\,pc\,cm$^{-3}$. An extragalactic origin is thus most plausible. 
\subsection{Galactic DM contribution}
We now investigate whether an intervening ionized region along the line of sight that is unaccounted for in electron density models could explain the observed DM excess, DM$_{\rm E}$. This ionized nebula could either be coincidentally aligned with the FRB, or one in which the burst progenitor is embedded. Taking the average between DM$_{\rm NE2001}$ and DM$_{\rm YMW16}$ as the maximum Galactic DM contribution along the line of sight, the unexplained amount of DM is DM$_{\rm E}$\,=\,155\,pc\,cm$^{-3}$. 

Thermal free-free emission from such a nebula could potentially be detected by surveys of the radio continuum. The primary parameter that determines the brightness of the radio emission from a homogeneous, isothermal \ion{H}{2} region is the emission measure, EM, which can be calculated from the DM,
\begin{equation}
    \rm{EM} = \rm{DM}^2\,\mathit{D}_{pc}^{-1} \,pc\, cm^{-6}, 
\end{equation}
where $D_{\rm pc}$ is the nebula diameter in units of pc.
Following the rationale of \cite{kon+14} and \cite{sshc+16}, we can estimate the free-free optical depth, $\tau_{\rm ff}$, from the EM,
\begin{equation}
    \tau_{\rm ff} = 4.4 \times 10^{-7} \, \rm{EM} \, \left(\frac{\mathit{T_e}}{8000\,\rm{K}}\right)^{-1.35}\,\left(\frac{\nu}{1\,\rm{GHz}}\right)^{-2.1},
\end{equation}
where $T_e$ is the electron temperature and $\nu$ is the frequency \citep{o61}. We require for the nebula to be optically thin (i.e., $\tau_{\rm ff} < 1$) to the lowest-frequency photons detected from FRB 20200125A ($\sim$\,300\,MHz). Assigning the entire excess DM to the nebula, we can set a lower limit on the optically thin nebula size. For DM$_{\rm E}$ = 155\,pc\,cm$^{-3}$, we obtain $D_{\rm pc }\,\gtrsim\,$ 0.13\,pc. Thus, the corresponding upper limit on EM is 1.81$\,\times\,10^5\,$pc\,cm$^{-6}$. 

Assuming a typical electron temperature of $T_e \approx 8000\,$K, it is clear that radio photon energies are much smaller than the average kinetic energy of electrons in the medium. Therefore, the flux density of the radio emission can be estimated from the Rayleigh-Jeans approximation of the Planck function at a brightness temperature $T_b$ and frequency $\nu$,
\begin{equation}
    S_\nu \approx \frac{2\,k_B\,T_b\,\nu^2}{c^2},
    \label{eq:flux}
\end{equation}
where $T_b$ depends on the opacity at frequency $\nu$,
\begin{equation}
    T_b = T_e\, (1 - e^{-\tau_\nu})
    \label{eq:Tb}
\end{equation}
\citep{rl79}. 


To identify the putative nebula in archival data, we must estimate the flux density by assuming a fiducial distance to the source. Along this line of sight, both NE2001 and YMW16 predict that the Milky Way electron density column starts plateauing at distances as small as 2.5\,kpc, and reaches 95\% of the Galactic maximum at 4\,kpc. It is therefore reasonable to assume that at 5\,kpc, the Milky Way contribution to the observed DM is maximal. We note that placing the source closer to the Solar System would increase the nebula's EM and the observed flux density, and consequently weaken the constraints. For a fiducial distance of 5\,kpc, the smallest angular extent on the sky is $5''$.


Based on the above relations, we evaluate whether the putative nebula would be detectable in the FIRST \citep{first-survey} and NVSS \citep{nvss} surveys conducted with the VLA telescope at 1.4 GHz, which have angular resolutions of $5.4''$ and $45''$, respectively. After calculating the brightness temperature $T_b$ at 1.4 GHz using equation (\ref{eq:Tb}) and multiplying equation (\ref{eq:flux}) by the sky area of the emitting region, we apply a correction factor to account for the reduction in flux density resulting from a source being either resolved or unresolved in the images. We find that an ionized region would be detectable above a 5-$\sigma$ level in FIRST data\footnote{\cite{first-survey} report a rms flux density for FIRST data of 0.15\,mJy.} if $D_{\rm pc }\,\lesssim\,1.8\,$pc. Without increasing the amount of ionized plasma, the flux density of a nebulae larger than 1.8\,pc at a distance of 5\,kpc would have a flux density below the detection threshold of the high-resolution FIRST data. Our calculations predict that the maximum flux density in the FIRST images is $\sim$\,10\,mJy, when the putative nebula is smallest in size (i.e., the case where the nebula angular size matches the resolution of the FIRST data). 
As for the lower-resolution NVSS data, we find that any source with $D_{\rm pc }\,\gtrsim\,0.3\,$pc would be detectable above a 5-$\sigma$ threshold\footnote{\cite{nvss} estimate the rms flux density in NVSS data at 0.45\,mJy.}. The flux density detected in the NVSS data would reach maximum value of $\sim$\,84\,mJy when the nebula has a size of 1.1\,pc. 

To identify viable candidates and eliminate contaminating sources such as radio stars, AGNs and other extragalactic sources, we adopt an approach similar than that presented by \cite{ckt99}, who compared the radio and mid-infrared flux densities to select ionized \ion{H}{2} regions such as planetary nebulae.  Our first step consists of selecting all catalogued objects from the ALLWISE catalog\footnote{Catalog built upon the Wide-field Infrared Survey Explorer (WISE) project \citep{wem+10} offering an enhanced sensitivity and improved astrometric precision. Data released in 2013 November available here: \url{http://wise2.ipac.caltech.edu/docs/release/allwise/}.} that have positions within our GBT beam.
We eliminate objects having a 25\,$\mu$m-flux density measured below a 5-$\sigma$ level and those having a 12-$\mu$m flux density stronger than their 25-$\mu$m flux density\footnote{We adopt a more flexible color-based criteria than \cite{ckt99}, where they required that the 12 to 25-$\mu$m flux density ratio be less than 0.35 \citep{pm88}.}. Most objects that were rejected were stars. Then, after visually examining the FIRST and NVSS images at the position of the remaining sources, we reject those for which radio emission is not detected in either images. This produced only one candidate that satisfied all criteria which is spatially coincident with 2MASX J14371326+0738174 \citep{tom+14}, the brightest galaxy in a cluster located at a redshift of 0.182. Thus, our analysis produces no potential Galactic ionized region.

In addition to radio emission, recombination radiation in the form of H$\alpha$ emission would be emitted from the putative nebula. To confirm the above conclusion independently, we calculate the expected H$\alpha$ flux from the EM associated with different nebular sizes in the Southern H$\alpha$ Sky Survey \citep[SHASSA;][]{gmr+01} following the methodology presented by \cite{kon+14}, for both a resolved and unresolved source. The scenario that predicts that lowest H$\alpha$ flux is one where the nebular size is minimized ($D_{\rm pc}=$ 0.13\,pc) and its distance maximized. Placing the nebula at a distance of 10\,kpc, the predicted H$\alpha$ flux is $\sim\,1.1\,\times\,10^{-13}$ erg\,s$^{-1}$\,cm$^{-2}$, above SHASSA's 5-$\sigma$ flux limit of $7.8\,\times\,10^{-14}$ erg\,s$^{-1}$\,cm$^{-2}$. The SHASSA data show no evidence of such emission region in our field, hence excluding the possibility of there being a small nebula below the CHIPASS detection limit. Additionally, we note that our beam is at a higher Galactic latitude than any large-scale Galactic structures visible in data from either SHASSA or the Wisconsin H$\alpha$ Mapper \citep[WHAM;][]{wham}. 



We note the rarity of Galactic ionized nebulae at such high latitudes ($b$ = 58.43$^\circ$). The HASH catalog of planetary nebulae \citep{hash} shows that there is only one confirmed source, planetary nebula G003.3+66.1 \citep{fbp13}, and two candidate nebulae located within 10$^\circ$ of the GBT beam center. These objects are located $> 7.0^{\circ}$ north of our beam, inconsistent with being associated with FRB 20200125A. 

Considering the above discussion and the fact that FRB 20200125A is clearly an outlier in terms of the DM distribution of the known pulsar population in this sky direction, we conclude that it is unlikely that the Milky Way is the contributing source of the anomalous amount of excess DM for FRB 20200125A. 

\subsection{Extragalactic association}
Electronic matter in the host galaxy of FRB 20200125A and the Milky Way halo would both contribute to the observed DM$_{\rm E}$. We can however calculate the maximum redshift allowed for the source assuming that the plasma excess is entirely provided by the intergalactic medium (IGM). Using the scaling between redshift and the IGM contribution to the DM (the Macquart relation; \citealt{mpm+20}), we obtain an upper limit on the redshift $z_{max}$ = 0.17, corresponding to a maximum comoving distance\footnote{Assuming the standard model of cosmology for an Euclidean Universe ($\Lambda \rm CDM$).} of $\sim$ 750\,Mpc. If we assume that the Galactic halo is contributing 50\,pc\,cm$^{-3}$ to the excess DM, the estimated redshift reduces to 0.12. 

Identifying a single galaxy as the host of FRB 20200125A is difficult due to 1) the large FWHM of the GBT beam (36$'$) and 2) the large range of redshifts allowed for the host galaxy. Based on the GLADE combined catalogue \citep{glade}, there are more than three dozen catalogued galaxies located within the beam that have redshifts in the allowed range. Moreover, GLADE
has a completeness $<\,50\%$ within 200\,Mpc \citep{glade}, leaving a substantial fraction of galaxies within 750\,Mpc uncatalogued. Given the diversity in host properties of localized FRBs (\citealt{clw+17, rcd+19, bdp+19, pmm+19, mnh+20}), the absence of repeat bursts and the lack of polarization information, isolating a single galaxy as the host of FRB 20200125A from our GBT detection alone is impossible. 

\section{All-Sky FRB Rate} \label{sec:rate}

The FRB rate at 350\,MHz can be estimated based on the detection of one event in $\sim$33,000 survey pointings amounting to a total on-sky time, $T$ = 45.5\,days. The average RFI masking fraction for these pointings is 20.7\%, 0.7\% of which is estimated to be in the time domain and has been subtracted from the aforementioned on-sky time. The remaining masking can be attributed to persistent RFI in the 360--380\,MHz frequency range at the telescope site resulting in a usable bandwidth, $\Delta \nu = 60$\,MHz, after accounting for roll-off at the bandpass edges.

We determine the effective solid angle of the beam, $\Omega_{\mathrm{eff}}$, using the approach proposed by \citet{jbm+19} which accounts for variation in the FRB rate across the beam FWHM due to the position-dependent beam response, $B(\Omega)$. The magnitude of this variation depends on the FRB source counts distribution, which is modelled as a power law with the number of FRBs having a flux density greater than $S$ scaling as $S^{-\gamma}$. The effective field of view is then defined as,
\begin{equation}
    \Omega_{\mathrm{eff}} (\gamma) = \int B(\Omega)^{1-\gamma} d\Omega,
\end{equation}
with the rate at boresight given by,
\begin{equation}
    R = \frac{N}{T\, \Omega_{\mathrm{eff}}}.
\end{equation}
Here $N$ is the number of FRBs detected by the survey and $\Omega_{\mathrm{eff}}$ = 0.27\,sq.\,deg. for the GBT beam assuming an Euclidean flux density distribution of FRBs ($\gamma = 1.5$). We find the on-axis rate to be $3.4^{+15.4}_{-3.3} \times 10^3$ FRBs sky$^{-1}$ day$^{-1}$ with the uncertainties representing the 95\% confidence interval for Poisson statistics.

\begin{figure*}[t]
    \gridline{\fig{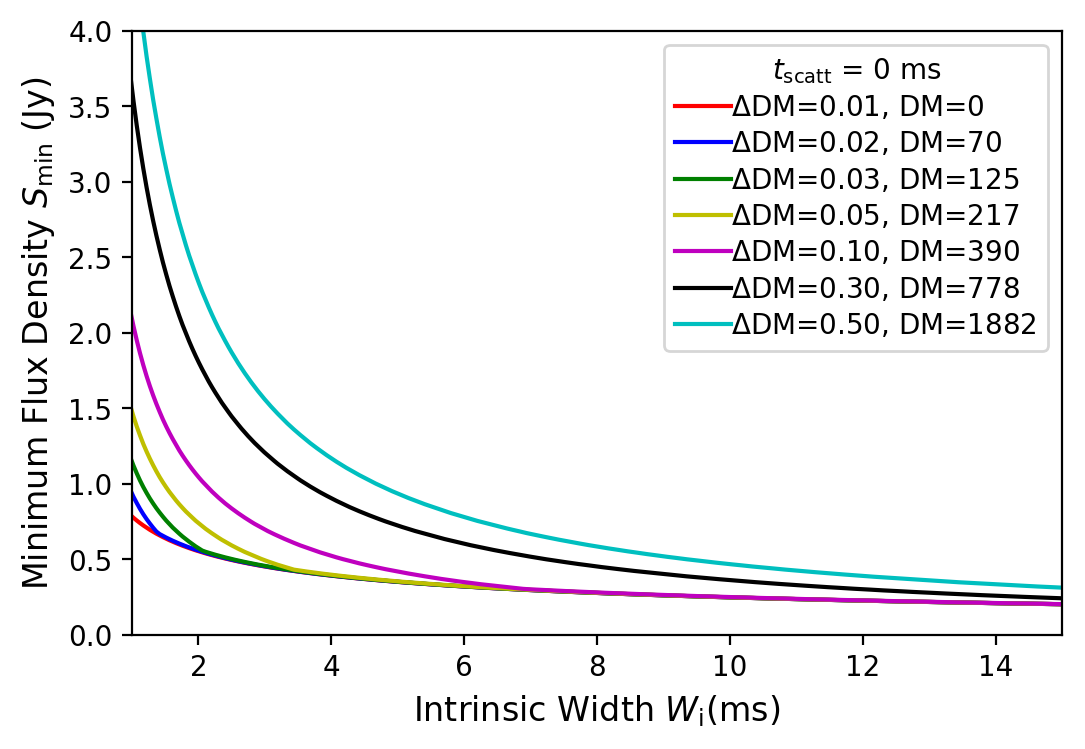}{0.5\textwidth}{(a)}
    \fig{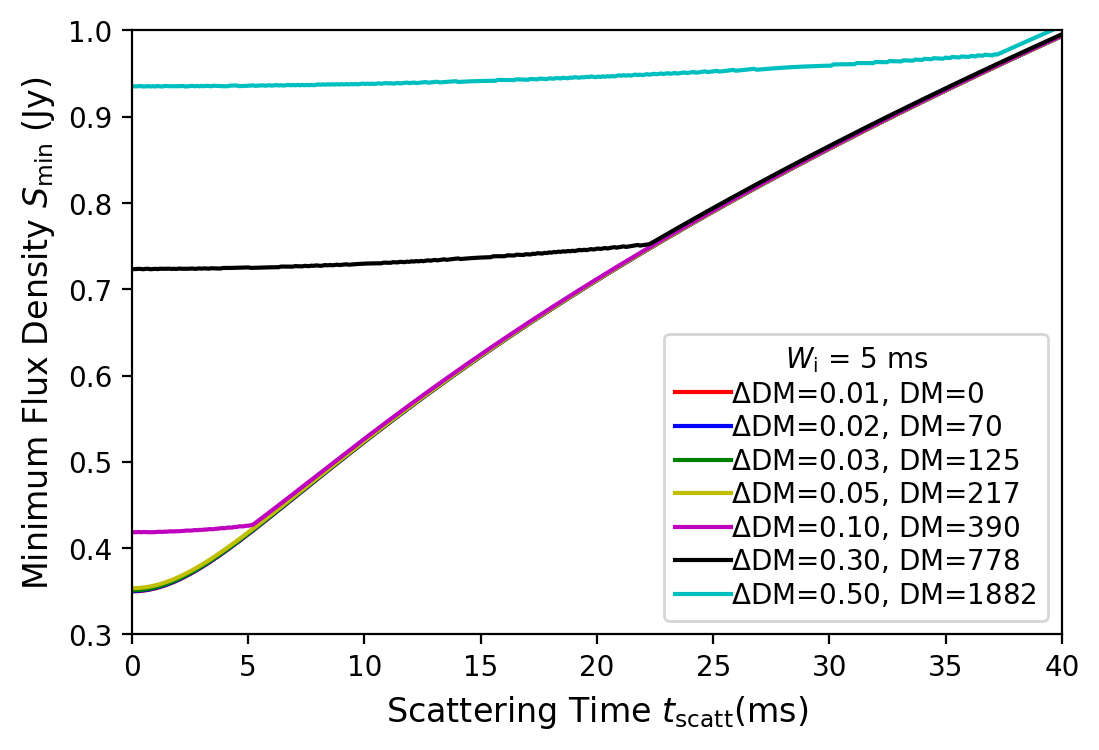}{0.5\textwidth}{(b)} }
    \caption{Threshold flux density, $S_\mathrm{min}$, plotted as a function of intrinsic pulse width $W_\mathrm{i}$ in panel (a) and scattering timescale $t_\mathrm{scatt}$ at 350\,MHz in panel (b). The threshold flux density varies with the DM step sizes used for the search. These step sizes are specified along with the corresponding lower bound of the trial DM range in units of pc\,cm$^{-3}$.}
    \label{fig:sensitivity}
\end{figure*}

The threshold flux density, $S_\mathrm{min}$, for detecting an FRB on-axis with our search pipeline can be determined using the radiometer equation, 
\begin{equation}
    S_\mathrm{min} = \frac{\beta\,S/N\, (T_\mathrm{rec}+T_\mathrm{sky})}{G\, W_\mathrm{i}} \sqrt{\frac{W_\mathrm{b}}{n_\mathrm{p} \Delta{\nu}}},
\end{equation}
where $T_\mathrm{sky}$ = 39\,K (median for all pointings included in the total on-sky time) and the observed pulse width, $W_\mathrm{b}$, is the quadrature sum of the intrinsic pulse width ($W_\mathrm{i}$), scattering timescale ($t_\mathrm{scatt}$), intrachannel dispersive smearing and sampling time \citep{cm03}. Values of other parameters are provided in Section \ref{sec:burst}. Although the S/N threshold for visual inspection of a candidate is 8 (see Section \ref{sec:cand}), the minimum detectable S/N could be higher as it depends on the intrinsic pulse width, burst DM and scattering timescale. This dependence arises due to the grouping and ranking code, \texttt{RRATtrap}, requiring detection at 20 DM trials for a pulse to be classified as a candidate \citep{ckj+17}. We obtain the minimum detectable S/N using the approach described in \citet{ckj+17} and find the threshold flux density to be 0.42\,Jy for $W_\mathrm{i} = 5$\,ms, $t_\mathrm{scatt} = 0$\,ms and a DM of 640\,pc\,cm$^{-3}$ (mean DM of known FRBs; \citealt{pbj+16}). However, the threshold flux density varies from 0.25\,Jy for a 10-ms duration burst to 2.0\,Jy for a burst of 1-ms duration (see Figure \ref{fig:sensitivity}). Additionally, the threshold is valid for FRBs with broadband emission. The sensitivity to narrow-band bursts, such as those emitted by repeating FRB sources \citep{hss+19, chime_rn1, chime_rn2}, is lower as our detection pipeline searches band-averaged time series for single pulses. In the absence of knowledge about the underlying distribution of FRB emission bandwidths, we do not modify our sensitivity threshold to account for narrow-band bursts, but note the quoted rate may be underestimated.
\begin{figure*}[t]
    \centering
    \includegraphics[width=0.55\textwidth]{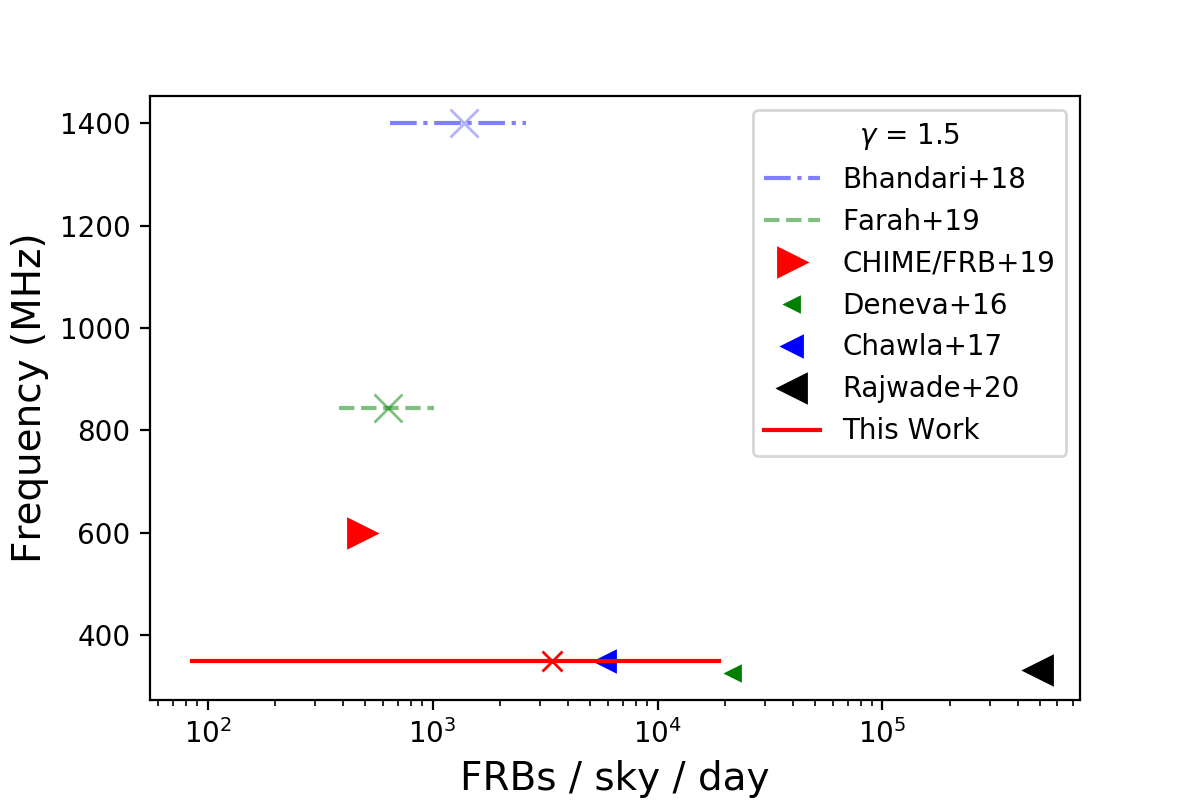}
    \caption{Rate measurements and limits for FRB surveys at different frequencies scaled to the flux density threshold for this work (0.25 Jy for an intrinsic pulse width of 10 ms) assuming an Euclidean source counts distribution ($\gamma$ = 1.5). The markers denote the published upper limits for the GBNCC survey \citep{ckj+17}, the AO327 survey \citep{dsm+16} and the search conducted by \citet{rms+20} using the Lovell telescope. The rate floor estimated by \citet{chime19} is also plotted here along with measurements from the SUPERB survey conducted with the Parkes telescope \citep{bkb+18} and the UTMOST project \citep{ffb+19}.}
    \label{fig:ratevsfreq}
\end{figure*}
Assuming an Euclidean flux density distribution of FRBs, we scale the upper limits previously reported by \cite{dsm+16} and \cite{rms+20} in the 300--400\,MHz frequency range to our search sensitivity, and we find that our measured rate is fully consistent with the scaled limits (see Figure \ref{fig:ratevsfreq}).

To allow for a meaningful comparison with the constraints on the FRB rate estimated by \citet{ckj+17}, we recompute their upper limit with the approach used in this work. We estimate it to be $<5.5 \times 10^3$ FRBs sky$^{-1}$ day$^{-1}$ above a flux density of 0.39\,Jy for an unscattered pulse with $W_\mathrm{i} = 5$\,ms. Although our S/N threshold for visual inspection was lower than that for the search conducted by \citet{ckj+17}, the lower median sky temperature ($T_\mathrm{sky}$ = 34\,K) and higher usable bandwidth ($\Delta \nu = 75$\,MHz) for their pointings resulted in the two searches having comparable sensitivities. 

We find that the mean rate derived from the detection of FRB 20200125A is consistent with the 95\% confidence upper limit reported by \citet{ckj+17}. We note, however, our upper bound on the confidence interval allows for rates larger than their upper limit. This tension in the maximum FRB-rate predictions for the GBNCC survey could be explained by the enhancement of our search sensitivity following a modification made to \texttt{RRATtrap}: we now allow the algorithm to search for associated detections in five neighbouring DM trials before classifying a single pulse event as noise. \citet{ckj+17} grouped detections only in the adjacent DM trials thus making their search more susceptible to RFI which can prevent detection of a pulse at certain DMs. Additionally, the use of the CyberSKA online portal for visual inspection in this work could potentially have resulted in a more rigorous search than the one conducted by \citet{ckj+17}. Nevertheless, our measured mean rate is consistent with \cite{ckj+17}, and we note that the errors on our measurement are large. Continued processing of GBNCC data with the enhanced search sensitivity will help increase the precision on our measurement.

We do not repeat the analysis performed by \citet{ckj+17} which constrained the mean spectral index of the FRB population by comparing the 1.4-GHz FRB rate with the rate at 350\,MHz. Their analysis relied on the assumption that the rate at 350\,MHz is equal to the 95\% confidence GBNCC upper limit implying that their spectral index constraints are still valid. Detection of more FRBs at these frequencies could reduce the Poissonian uncertainties associated with the rate measurement and allow for stricter constraints on the ensemble spectrum of FRBs.


\section{Conclusion} \label{sec:conclu}
In this work, we have presented FRB 20200125A, the first FRB discovered by the GBNCC survey and also the first source to be discovered by a large-scale survey for radio pulsars and fast transients operating at observing frequencies below 400\,MHz. The burst was detected in a high-latitude ($b=58.43^\circ$) pointing, and shows no evidence of interstellar scattering. While FRB 20200125A was detected at a marginal S/N of 8.1, all single-pulse candidates generated by our processing pipeline having a S/N$>$8 were subsequently confirmed as RRATs. Consequently, we consider FRB 20200125A highly likely to be astrophysical. 

Along this line of sight, electron density models NE2001 and YMW16 predict nearly identical maximum Galactic DMs that are more than seven times smaller than the source DM. Moreover, known field and globular-cluster pulsars located within 25$^\circ$ of the pointing position all have DM values consistent with a Galactic origin. We examined multi-wavelength archival data to identify a potential ionized region that could explain the anomalous amount of DM. Our analysis rules out the existence of such nebulae in the vicinity of FRB 20200125A. Based on these findings, we conclude that the source is extragalactic. 

We computed the all-sky FRB rate at 350\,MHz based on the detection of FRB 20200125A, the only FRB candidate in all 33,000 GBNCC survey pointing searched with the upgraded pipeline. We calculated a rate of $3.4^{+15.4}_{-3.3} \times 10^3$ FRBs sky$^{-1}$ day$^{-1}$ above a peak flux density of 0.42\,Jy for a unscattered pulse having an intrinsic pulse width of 5\,ms, assuming an Euclidean flux density distribution of FRBs. \cite{ckj+17} reported an upper limit of $3.6 \times 10^3$ FRBs sky$^{-1}$ day$^{-1}$ for a similar pulse above a peak flux density of 0.63\,Jy. We recompute their rate with the approach presented in this work, and obtain an 
upper limit of $5.5 \times 10^3$ FRBs sky$^{-1}$ day$^{-1}$ above a flux density of 0.39\,Jy. The upper bound of rate confidence interval we presented in this work is inconsistent with these upper limits. We suspect that this tension arises from the enhancement in search sensitivity resulting from our pipeline upgrades, notably the modifications made to the grouping algorithm and the more rigorous candidate visual inspection.

Our detection of FRB 20200125A, as well as previous observations of repeat bursts from FRB 180916.J0158+65 by \cite{cab+20} and \cite{pbp+20}, demonstrate that any frequency cutoffs invoked in bursting mechanisms for FRBs must be occur below 300\,MHz. Moreover, the local environment of at least some FRBs must be optically thin to free-free absorption. Similarly to the observations reported by \cite{cab+20} and \cite{pbp+20}, our strict constraint on the observable scattering for the source, $< 655\,$ms, at 350\,MHz also suggests that the circumburst environment of FRB 200125A does not have strong scattering properties. 

We plan follow-up observations of FRB 20200125A to search for repeat bursts which would remove any ambiguity regarding its cosmic nature, and potentially enable an interferometric localization. We also continue our search for FRBs in new GBNCC pointings, and plan on reprocessing pointings that were searched with earlier versions of the pipeline. 

\section*{Acknowledgements}
The Green Bank Observatory is a facility of the National Science Foundation (NSF) operated under cooperative agreement by Associated Universities, Inc. The CyberSKA project was funded by a CANARIE NEP-2 grant. Computations were made on the supercomputers Guillimin and B\'{e}luga in Montr\'{e}al, managed by Calcul Qu\'{e}bec and Compute Canada. The operation of this supercomputer is funded by the Canada Foundation for Innovation (CFI), NanoQu\'{e}bec, RMGA and the Fonds de recherche du Qu\'{e}bec$-$Nature et technologies (FRQ-NT). \\
EP is a Vanier Canada Graduate Scholar. PC is a recipient of the FRQ-NT Doctoral Research Award. VMK acknowledges the NSERC Discovery Grant, the Herzberg Award, FRQ-NT and the Centre de recherches en astrophysique du Qu\'{e}bec (CRAQ), the Canadian Institute For Advanced Research (CIFAR) and the Webster Foundation Fellowship, a Distinguished James McGill Chair and the Trottier Chair in Astrophysics and Cosmology. JvL acknowledges funding from the European Research Council under the European Union’s Seventh Framework Programme (FP/2007-2013) / ERC Grant Agreement n. 617199 (`ALERT'), and from Vici research programme `ARGO' with project number 639.043.815, financed by the Dutch Research Council (NWO). Pulsar research at UBC is supported by an NSERC Discovery Grant and by CIFAR. DLK, JKS, MD, MM, SMR and XS acknowledge the NANOGrav Physics Frontiers Center, which is supported by the NSF award 1430284. EFL, MM and MS acknowledge the NSF OIA award number 1458952. IHS and SMR are CIFAR Senior Fellows.

\bibliography{main.bib}{}
\bibliographystyle{aasjournal}
\end{document}